\documentclass[a4paper,12pt]{article}

\usepackage[usenames]{color}
\usepackage{amssymb}
\usepackage{amsmath}
\newcommand{\half}{\frac{1}{2}}
\newcommand{\del}{\partial}
\newcommand{\ep}{\epsilon}
\newcommand{\bra}[1]{\langle #1 |}
\newcommand{\ket}[1]{| #1 \rangle}

\newcommand{\nn}{\nonumber}
\newcommand{\p}{\partial}
\newcommand{\pb}{\bar \partial}
\newcommand{\ti}{\tilde}
\newcommand{\al}{\alpha}
\newcommand{\be}{\beta}
\newcommand{\ga}{\gamma}
\newcommand{\de}{\delta}
\newcommand{\si}{\sigma}
\newcommand{\la}{\lambda}

\newcommand{\vev}[1]{\langle #1 \rangle}
\newcommand{\Vev}[1]{\left\langle #1 \right\rangle}
\newcommand{\mb}{\mathbf}
\newcommand{\nt}{\notag\\}
\newcommand{\V}{\mathcal{V}}
\newcommand{\m}{\mathrm{m}}
\newcommand{\eFT}{e^{\int\frac{d\si}{2\pi}FT}}
\newcommand{\eTT}{e^{-\frac14\int\frac{d\si}{2\pi}T^2}}
\newcommand{\db}{\displaybreak[0]}
\newcommand{\ap}{\alpha}

\newcommand{\Sym}{{\rm Sym}}
\newcommand{\mV}{2i\{b_0^-,O\}}

\newcommand{\bulk}{{\rm bulk}}

\newcommand{\ev}[1]{\left \langle #1 \right \rangle}
\newcommand{\newvac}{\ket{Y\tilde Y}}
\newcommand{\bs}{\bra{B}}
\newcommand{\bea}{\begin{align}}
\newcommand{\eea}{\end{align}}

\begin{document}

\begin{titlepage}
\title{\vspace{1cm}
On the general action of\\ boundary (super)string field theory}
\author{\vspace{2mm}
Akira Ishida$^1$\thanks{E-mail: \ttfamily ishida@skku.edu}~~and
Shunsuke Teraguchi$^2$\thanks{E-mail: \ttfamily teraguch@eken.phys.nagoya-u.ac.jp}\\[4mm]
$^1$\textit{\normalsize Department of Physics, BK21 Physics Research Division,}\\
\textit{\normalsize and Institute of Basic Science, Sungkyunkwan University,}\\
\textit{\normalsize Suwon 440-746, Korea}\\[3mm]
$^2$\textit{\normalsize Department of Physics, Nagoya University,}\\
\textit{\normalsize Nagoya 464-8602, Japan}
}
\date{}
\maketitle
\thispagestyle{empty}
\vspace{-110mm}
\begin{flushright}
arXiv:0805.1826\\
May, 2008
\end{flushright}
\vspace{100mm}
\begin{abstract}
We reconstruct boundary superstring field theory via boundary states.
After a minor modification of the fermionic two-form,
all the equations needed for Batalin-Vilkovisky formulation are simply
represented by closed string oscillators
and the proof of gauge invariance
is drastically simplified.
The general form of the action of boundary superstring field
theory is also obtained without any assumption
and found to take exactly the same form as the bosonic one.
As a special case of this action, we revisit the conjecture
that the action is simply given by the disk partition function
when matter and ghosts are completely decoupled.
\end{abstract}

\end{titlepage}

\section{Introduction}
Boundary string field theory (BSFT) \cite{Witten:1992qy,Witten:1992cr}
is one of off-shell formulations
of string theory. Though it is a version of covariant open string field theory,
the formulation is close to worldsheet sigma-model rather than
the other open string field theories \cite{Witten:1985cc,Berkovits:1995ab}.
BSFT has been applied to unstable D-brane systems including tachyons
(for a review, see \cite{Sen:2004nf}) and it turned out to be successful in
describing such systems.
For example, it gives the exact form of the tachyon potential
\cite{Gerasimov:2000zp,KMM,KMM2}.

BSFT is formulated on the space of all boundary interactions specified
by couplings $\la$'s in worldsheet sigma-models based on the
Batalin-Vilkovisky (BV) formalism.
In bosonic string theory, the spacetime action $S$ of BSFT,
which is a function of $\la$'s, is defined through the equation,
\begin{equation}
dS=\frac12\int_0^{2\pi}\!d\si d\si'\,
\vev{d\mathcal{O}(\si)\{Q_B,\mathcal{O}(\si')\}}_\lambda,\label{dS}
\end{equation}
where the correlator $\vev{\cdots}_\la$ is evaluated in a worldsheet
sigma-model on a disk with the boundary perturbation defined by $\la$'s.
Here, 
$\mathcal{O}(\si)$ is a boundary operator which is also specified by $\la$'s,
and $\si$ parametrizes the boundary of the disk.
$Q_B$ is the bulk BRST operator.
The gauge invariance of this action is guaranteed by
the BV formalism.
Under the assumption of decoupling of matter and ghosts,
it was shown in \cite{Witten:1992cr,Shatashvili:1993ps} that the action $S$ is
related to the disk partition function $Z$ and $\be$-functions of the
worldsheet sigma-model as
\begin{equation}
S(\la)=\left(-\be^i(\la)\frac{\p}{\p\la^i}+1\right)Z(\la).
\end{equation}

A proposal for boundary superstring field theory (super BSFT) for non-BPS
D-branes was first made in \cite{KMM2}, where the action of super BSFT is
rather phenomenologically identified with the
corresponding disk partition function:\footnote{
This was originally discussed for massless modes in \cite{oldS=Z}.
The extension to D-brane-anti-D-brane system was given in \cite{DDbar}.}
\begin{equation}
S(\la)=Z(\la).\label{S=Z}
\end{equation}
Soon later, justification for this proposal was attempted by
\cite{Marino,Niarchos:2001si} based on the BV formalism.
In order to show the conjectured relation between the action and the partition function,
the following trick given in \cite{Witten:1992cr} has been used.
First, assume that the matter system consists of two decoupled subsystems.
Then, by using the local integrability of the definitional equation \eqref{dS},
one can express the action of BSFT by partition functions and
several one-point functions of the two subsystems.
In the case of superstrings, it is plausible that the
one-point functions above, which are those of fermionic operators, vanish and
the action simply reduces to the partition function of the whole system.
Thus, the conjecture has been indirectly shown.

Recently, bosonic BSFT was reformulated in terms of the closed string Hilbert
space \cite{Teraguchi:2006tb}. The main advantage of this formulation
is that we can obtain the action $S$ itself
by a simple algebraic calculation without any assumption:
\begin{equation}
S=\frac{1}{4}\bra{B}e^{2i\{b_0^-,O\}}c^-_0Q_Bc_0^-\ket{0}-\frac{i}{2} \bra{B}\Sym\big[e^{2i\{b_0^-,O\}};\{Q_B,O\}\big]c_0^-\ket{0}.
\label{ourBSFTaction}
\end{equation}
Then, it is reasonable to expect that
this formulation also works for superstrings.
The main interest of this paper is to reconsider the construction of super BSFT 
in the same manner as in \cite{Teraguchi:2006tb} and obtain the general action $S$ for superstrings without any assumption.
For this purpose, we need some modification for the fermionic two-form, which is one of the key ingredients for BV formulation of BSFT.
Under this new definition,
the fermionic two-form is much more simply represented in the closed string Hilbert space
without relying on bosonized superconformal ghosts.
Furthermore, in this formulation, 
the proof of the gauge invariance of the action is completely analogous to the one for bosonic string,
and greatly simplified compared to the one given in \cite{Niarchos:2001si}.
The general action of super BSFT is also obtained without any assumption 
and it takes exactly the same form as the bosonic one.
Based on this general action, we revisit the conjectured relation, $S=Z$.
We also argue some common features on bosonic and super BSFT.

The organization of this paper is as follows. In section \ref{revBSFT},
we give a short review on the previous construction of BV formulation for super BSFT
and propose a new definition of the fermionic two-form.
We construct super BSFT in the closed string Hilbert
space and evaluate the action in section \ref{reformBSFT}. 
The general form \eqref{act} of super BSFT action
is one of our main results of this paper.
Two formal aspects of the BSFT action, the expansion form and the gauge transformation, shared among bosonic and super BSFT, are provided in section \ref{expand}.
In section \ref{revisit}, we reconsider 
the well-known relation $S=Z$ in our formulation.
The final section is devoted to summary and discussion.
Our convention and some explicit calculations concerning the boundary fermion for non-BPS systems are given in appendices.

\section{BV formulation of super BSFT and its modification}\label{revBSFT}
In this section, we shall briefly review the construction of super BSFT
by mostly following \cite{Niarchos:2001si}, where boundary operators in 0-picture
are regarded as fundamental.
To keep the worldsheet supersymmetry, it is convenient to use the superfield formalism.
The bulk worldsheet action of NSR superstring in the superfield formalism is compactly given by\footnote{
Throughout this paper, we use the convention $\al'=2$.}
\begin{equation}
S_\bulk=\frac{1}{4\pi}\int d^2zd^2\theta D_{\bar{\theta}}{\bf X}^\mu D_{{\theta}}{\bf X}_\mu
+\frac{1}{2\pi}\int d^2zd^2\theta BD_{\bar{\theta}}C
+\frac{1}{2\pi}\int d^2zd^2\theta \ti BD_\theta\ti C,
\label{wsaction} 
\end{equation}
where the worldsheet superfields ${\bf X}$, $B$ and $C$ are written by
the usual worldsheet fields,
\begin{align}
{\bf X}^\mu({\bf z},{\bf \bar z})&=X^\mu+i\theta\psi^\mu+i\bar\theta\tilde\psi^\mu+\theta\bar\theta F^\mu,\label{superX}\\
B({\bf z})&=\beta(z)+\theta b(z),\\
C({\bf z})&=c(z)+\theta \gamma(z),
\end{align}
and $\ti B$ and $\ti C$ are the anti-holomorphic counterparts of $B$ and $C$.
Here, $\theta$ and $\bar\theta$ are fermionic coordinates and the superderivatives are given by
\begin{equation}
D_{{\theta}}=\del_\theta+\theta\del,\qquad
D_{\bar{\theta}}=\del_{\bar\theta}+\bar\theta\bar\del.
\end{equation}
One can consider a boundary perturbation which keeps the worldsheet supersymmetry by introducing a boundary action of the form
\begin{equation}
S_{\rm bdy}=\int \frac{d\sigma d\theta}{2\pi} {\cal V}(\sigma,\theta).
\end{equation}
Here, ${\cal V}(\sigma,\theta)$ is a boundary perturbation
written by superfields with 0-picture and ghost number 0.
This boundary perturbation can be expanded by boundary couplings $\lambda$'s,
\begin{equation}
\mathcal{V}=\sum_I \lambda^I \mathcal{V}_I,
\end{equation}
where $\mathcal{V}_I$ is a basis of boundary operators.
In the formulation of \cite{Niarchos:2001si}, the boundary operator ${\cal O}$,
which is related to the above boundary perturbation by
${\cal V}=b^{\rm BSFT}_{-1}{\cal O}$,
is considered as the basic object of this string field theory.
The operator
$b^{\rm BSFT}_{-1}$ has ghost number $-1$ and its precise definition
is given in \cite{Witten:1992qy}.
Thus, ${\cal O}$ has picture number 0 and ghost number 1.

In order to construct a boundary string field theory, we need a fermionic vector $V$ and a fermionic two-form $\omega$
in the space of all boundary interactions, which satisfy the following three properties:
\begin{align}
V^2=0:&\quad\rm{nilpotency},\label{nilpotency}\\
d\omega=0:&\quad\rm{closedness},\label{closedness}\\
d(i_V\omega)=0:&\quad V\rm{-invariance}.\label{V-invariance}
\end{align}
The natural choice of the fermionic vector $V$ is the one generated by the bulk BRST operator $Q_B$.
The nilpotency of $V$ immediately follows from the nilpotency of $Q_B$.
The two-form $\omega$ is defined by two-point correlation functions of the deformed worldsheet theory.
In the case of super BSFT, the choice of the two-form is more subtle due to the complexity of the notion of picture \cite{FMS}.
In \cite{Niarchos:2001si}, the two-form $\omega$ was defined in the following way:
\begin{equation}
\omega(\delta_1{\cal O},\delta_2{\cal O})=(-)^{\ep(\delta_1 {\cal O})}\int d\sigma_1d\sigma_2d\theta_1d\theta_2
\ev{Y(\sigma_1)\delta_1{\cal O}(\sigma_1,\theta_1)Y(\sigma_2)\delta_2{\cal O}(\sigma_2,\theta_2)}_\la.
\end{equation}
Here, $Y(\sigma)$ is the inverse picture-changing operator,
\begin{equation}
Y=c \del\xi e^{-2\phi},
\end{equation}
and $\phi$ is the boson used to bosonize superconformal ghosts,\footnote{
We regard $e^{q\phi}$ for odd $q$ as fermionic so that
$\be$ and $\ga$ obey the correct statistics.}
\begin{equation}
\be=e^{-\phi}\p\xi,\qquad \ga=\eta e^{\phi},
\end{equation}
where $\xi$ and $\eta$ are fermions of dimension zero and one, respectively.
In the above definition of $\omega$, an inverse picture-changing operator is inserted\footnote
{In \cite{Marino}, the two-form is directly defined by inserting operators of picture number $-1$ without explicitly using the superfield formalism or the inverse picture-changing operators.} at the same position of each of the two boundary operators ${\cal O}(\sigma,\theta)$ in order to decrease the picture number of ${\cal O}$ by one and saturate picture number $-2$.

However, in this paper, we propose a modified definition of the two-form $\omega$ instead of the above one.
We find the following new definition much more convenient for our purpose of reformulating super BSFT in terms of boundary states.
We define the two-form $\omega$ by inserting a double-step inverse picture-changing operator $Y\tilde Y$ at the center of the disk,
$z=e^{\tau-i\si}=0$, instead of the boundary:
\begin{equation}
\omega(\delta_1{\cal O},\delta_2{\cal O})=(-)^{\ep(\delta_1 {\cal O})}\int d\sigma_1d\sigma_2d\theta_1d\theta_2
\ev{Y\tilde Y(0)\delta_1{\cal O}(\sigma_1,\theta_1)\delta_2{\cal O}(\sigma_2,\theta_2)}_\la.
\label{our newdef omega}
\end{equation}
Here the double-step inverse picture-changing operator is nothing but the product of the holomorphic inverse picture-changing operator and the anti-holomorphic one.
This type of picture-changing operator was first used in \cite{DSTEP}
in order to overcome the singular behavior of cubic superstring field theory
\cite{SuperCSFT} due to the collision of picture-changing operators
at the midpoint.

By using these fermionic vector and two-form, (modified) super BSFT action is defined by the equation
\begin{equation}
dS_{\rm BSFT}=(-)^{\ep(\delta_1 {\cal O})}\int d\sigma_1d\sigma_2d\theta_1d\theta_2
\ev{Y\tilde Y(0)d{\cal O}(\sigma_1,\theta_1)\{Q_B,{\cal O}(\sigma_2,\theta_2)\}}_\la.
\end{equation}
One may think that positions of picture-changing operators do not matter anyway. However, in superstring field theory where off-shell operators should be considered as well, we cannot freely change the positions of picture-changing operators.
Thus, our action of super BSFT considered here is, in principle, different from the ones in \cite{Marino,Niarchos:2001si}.
The advantage of this modified version of the two-form $\omega$ is that,
by noting that
the double-step inverse picture-changing operator commutes with
$Q_B$ and $b^{\rm BSFT}_{-1}$,
\begin{equation}
\{Q_B, Y\tilde Y\}=0,\qquad \{b^{\rm BSFT}_{-1}, Y\tilde Y(0)\}=0,\label{propYY}
\end{equation}
one can avoid all the complexity and the subtleties argued in the appendix A in $\cite{Niarchos:2001si}$,
related to inverse picture-changing operators.
Therefore, as we will see in the next section, it is straightforward\footnote{In the large Hilbert description, we must insert one additional $\xi$ in order to saturate the zero-mode. One may wonder whether the insertion of $\xi$ makes any trouble. However, if BRST operator hits the $\xi$, the term does not contain $\xi$ zero-mode and just vanishes. So it does not affect the argument.} to apply the previous proof \cite{Witten:1992qy,Shatashvili:1993ps,Teraguchi:2006tb} for bosonic BSFT to the superstring case under this new definition.

\section{General form of super BSFT action}
\label{reformBSFT}
In \cite{Teraguchi:2006tb}, bosonic BSFT was reformulated in terms of boundary states.
In that formulation, one can perform the integration of the definitional equation \eqref{dS} without making any assumption or approximation and obtain the general form of the action itself.
We proceed to formulate the above super BSFT along the same lines.\footnote{
In \cite{Kmtx},
the matter sector of super BSFT has been discussed in detail by using the boundary state formalism.}

First note that the insertion of the double-step inverse picture-changing operator $Y\tilde Y$ at the center of the disk corresponds to considering the following state in the closed string Hilbert space:
\begin{align}
Y(0)\tilde Y(0)&\sim\lim_{z\to0}Y(z)\ti Y(\bar z)\ket0\nt
&=\lim_{z\to 0} \left(zc(z)\p\xi(z) e^{-\phi(z)}
e^{-\phi(0)}\right)\left(\bar z \ti c(\bar z)\pb\ti\xi(\bar z)
e^{-\ti\phi(\bar z)}e^{-\ti\phi(0)}\right)\ket{0}\nt
&=-\lim_{z\to 0} (z\be(z))(\bar z \ti\be(\bar z))c(z)\ti c(\bar z)\ket0_{-1,-1}
=-\beta_{-1/2}\tilde\beta_{-1/2}\ket{\Omega}\equiv \newvac.\label{newvac}
\end{align}
Here $\ket0_{-1,-1}\equiv e^{-\phi(0)-\ti\phi(0)}\ket0$ represents the closed string vacuum with picture number $(-1,-1)$ and
$\ket{\Omega}=c_1\tilde c_1\ket0_{-1,-1}$ is the Fock vacuum for closed string oscillators.
Remarkably, once we introduce this state $\newvac$ to calculate the correlation functions,
there is neither complexity nor subtleties regarding the bosonization
of superconformal ghosts and the picture-changing operation,
since all the boundary operators we consider here do not have any picture.
It immediately follows from \eqref{propYY} that the state $\newvac$ is annihilated by the BRST operator and
$b_0^-\equiv (b_0-\ti b_0)/2$, which comes from $b_{-1}^{\rm BSFT}$
in the boundary action:
\begin{equation}
Q_B\newvac=0, \qquad b_0^-\newvac=0.
\end{equation}
In addition, $\newvac$ has the total ghost number 0.
Thus, $\newvac$ shares common properties with the $SL(2,C)$ vacuum $\ket{0}$ in bosonic string.

Making use of this state, we propose the following definitions of the nilpotent vector $V$ and the closed two-form $\omega$,
\begin{align}
\delta_V O&\equiv\{Q_B,O\}, \label{defofV}\\
\omega&\equiv\half\bs\Sym[e^{\mV};dO,dO]\ket{Y\tilde Y}.
\label{defofomega}
\end{align}
Here, $O$ represents a boundary operator ${\cal O}(\sigma,\theta)$ after $\sigma$ and $\theta$ integrations,
\begin{equation}
O\equiv\int_0^{2\pi}\frac{d\sigma d \theta}{2\pi} {\cal O}(\sigma,\theta),
\end{equation}
which can be expanded by couplings $\lambda^I$ and basis $\mathcal{O}_I$ as $O=\sum_I\lambda^IO_I$.
In the construction of BSFT, we always define statistics of worldsheet couplings, namely string fields, so that this $O$ after $\theta$ integration is fermionic.
The symbol $\Sym[\cdots]$ is defined in \cite{Teraguchi:2006tb}:
\begin{align}
&\Sym[e^{-V};O_1,O_2,\cdots,O_n]\nn\\
&=\int_0^1dt_1 \int_{t_1}^1dt_2 \cdots \int_{t_{n-1}}^1 dt_n  e^{-t_1V}O_1
e^{-(t_2-t_1)V}O_2\cdots O_n e^{-(1-t_n)V}
\pm({\rm perms}).
\end{align}
$\bs$ can be any on-shell boundary state satisfying $\bs Q_B=0$ and $\bs b_0^-=0$, depending on
the open string vacuum at issue.
In superstring theory, we have extra degrees of freedom to specify the sign of boundary conditions for fermionic variables.
The physical boundary state which survives after GSO projection should be some linear combination of them.
Since, in our formulation of super BSFT, the two-form (and the action as well) is comprised of an inner product of such boundary states with the GSO-even closed string state $\ket{Y\tilde Y}={\cal P}_{\rm GSO}\ket{Y\tilde Y}$, the bra state is automatically chosen to be a GSO-even combination of boundary states.

Now we shall show that our vector and two-form really satisfy the properties
(\ref{nilpotency})-(\ref{V-invariance}).
Fortunately, the algebraic proof given in \cite{Teraguchi:2006tb} is directly applicable to the current case by replacing the bosonic closed string vacuum $\ket{0}$ with $\ket{Y\tilde Y}$, both of which are ghost number 0 and annihilated by $b_0^-$ and (the corresponding) BRST operators.
In the following, we repeat the argument given in \cite{Teraguchi:2006tb} to make this paper self-contained.
A more detailed explanation can be found in the reference.
The nilpotency of the vector $V$ simply follows from the nilpotency
of $Q_B$\footnote{
For a non-BPS D-brane, the auxiliary boundary superfield
$\mb\Gamma$ is introduced \cite{bdryfermion} to take the GSO-odd sector into account.
In this case, we should modify the BRST operator so that the anti-commutator
$\{b_0^-,Q_B\}$ generates the rotation in 
this sector as well. This modification is necessary to prove the $V$-invariance of the two-form $\omega$.
We discuss the BRST transformation of this auxiliary boundary superfield in Appendix \ref{abs}.
} as usual.
The closedness of $\omega$ comes from the fact that the state $\ket{Y\tilde Y}$ is annihilated by $b_0^-$:
\begin{align}
d\omega&=i\bs\Sym\big[e^{\mV};\{b_0^-,dO\},dO,dO\big]\newvac\nt
&=\frac{i}3\bs\Sym\big[e^{\mV};dO,dO,dO\big]b_0^-\newvac=0.
\end{align}
Note that this proof of the closedness of $\omega$ is much simpler and transparent than the one attempted in \cite{Niarchos:2001si}, where BRST invariance of unperturbed correlators is further employed.
The invariance of $\omega$ under the transformation generated by $V$ is also proved without any difficulty.
\begin{align}
d(i_V\omega)&=2i\bs\Sym\big[e^{\mV};\{b_0^-,dO\},dO,\{Q_B,O\}\big]\newvac\nt
&\qquad-\bs\Sym\big[e^{\mV};dO,\{Q_B,dO\}\big]\newvac\nn\\
&=-i\bs\Sym\big[e^{\mV};dO,dO,\big[b_0^-,\{Q_B,O\}\big]\big]\newvac\nt
&\qquad-i\bs\Sym\big[e^{\mV};dO,dO,[Q_B,\{b_0^-,O\}]\big]\newvac\nn\\
&=-\frac{i}{2}\bs\Sym\big[e^{\mV};dO,dO,[L_0-\tilde{L}_0,O]\big]\newvac=0.
\end{align}
In the last line, we have used the rotational symmetry generated by the operator $L_0-\tilde{L}_0$ as usual.
In the above proof, we have used the following properties:
\begin{equation}
\bs b_0^-=\bs Q_B=b_0^-\newvac=Q_B\newvac=0.\label{propofvac}
\end{equation}
These are nothing but the properties used in the proof \cite{Teraguchi:2006tb} for bosonic BSFT:
\begin{equation}
\bra{N}b_0^-=\bra{N}Q_B=b_0^-\ket{0}=Q_B\ket{0}=0.
\end{equation}
Thus, our modified two-form (\ref{defofomega}) naturally satisfies the desired property (\ref{closedness}) and (\ref{V-invariance}).
Therefore, we can define a gauge invariant action $S$ of super BSFT using these ingredients:
\begin{equation}
dS=\bs\Sym\big[e^{\mV};dO, \{Q_B,O\}\big]\ket{Y\tilde Y}.\label{defofS}
\end{equation}

In \cite{Teraguchi:2006tb}, the most important advantage of rewriting BSFT by boundary states is
that the equation defining the action can be easily integrated by performing simple algebraic operations without any assumption or approximation.
Therefore, we expect that the above equation can also be integrated in the same manner.
Again, formally, the calculation is completely the same as the bosonic one \cite{Teraguchi:2006tb} and we have
\begin{equation}
dS=d\Bigg(
\frac{1}{4}\bs e^{\mV}c^-_0Q_Bc_0^-\newvac-\frac{i}{2} \bs\Sym\big[e^{\mV};\{Q_B,O\}\big]c_0^-\newvac\Bigg),
\end{equation}
which leads to the action itself
\begin{equation}
S=\frac{1}{4}\bs e^{\mV}c^-_0Q_Bc_0^-\newvac-\frac{i}{2} \bs\Sym\big[
e^{\mV};\{Q_B,O\}\big]c_0^-\newvac.\label{act}
\end{equation}
Thus, the action of super BSFT formally takes exactly the same form as the bosonic one \eqref{ourBSFTaction}.

\section{Expansion form and gauge transformation of BSFT}\label{expand}
As we have seen in the previous section, the formal expression of BSFT action is the same between bosonic and supersymmetric one. In this section, we develop some formal aspects of the BSFT action.
From now on, we use the expression for bosonic BSFT \eqref{ourBSFTaction} to describe the general action of BSFT for simplicity.
However, in the following, one can always recover the expressions for super BSFT by simply replacing $\ket{0}$ with $\ket{Y\tilde Y}$ and reinterpreting all the quantities by their supersymmetric cousins.
Expanding the general action \eqref{ourBSFTaction} in terms of string field $O$, we have
\begin{align}
S&=\frac{1}{4}\bra{B}c^-_0Q_Bc_0^-\ket{0}\nt
&\quad+\frac{1}{4}\sum_{n=0}^\infty\frac{(2i)^{n+1}}{(n+1)!}\bra{B}\Big[
\{b_0^-,O\}^{n+1}c_0^-Q_Bc_0^-
-\sum_{m=0}^{n}\{b_0^-,O\}^{n-m}\{Q_B,O\}\{b_0^-,O\}^mc_0^-\Big]\ket{0}.
\end{align}
One can evaluate the second term of the above expression order by order and
find that it takes rather simple form:
\begin{align}
S&=\frac{1}{4}\bra{B}c^-_0Q_Bc_0^-\ket{0}
+\sum_{n=0}^\infty\frac{(2i)^{n}}{(n+2)!}\sum_{m=0}^{n}\bra{B}(Ob_0^-)^{n-m}OQ_BO(b_0^-O)^m\ket{0}\nn\\
&=\frac{1}{4}\bra{B}c^-_0Q_Bc_0^-\ket{0}+\frac{1}{2}\bra{B}OQ_BO\ket{0}+\frac{i}{3}\bra{B}\big(OQ_BOb_0^-O+Ob_0^-OQ_BO\big)\ket{0}
+\cdots\,.
\end{align}
In order to interpret this action, it would be convenient to make a field redefinition, $O\rightarrow g_o O$, and an overall scaling, $S\rightarrow \frac{1}{g_o^2}S$, so that we have
\begin{align}
S&=\frac{1}{4g_o^2}\bra{B}c^-_0Q_Bc_0^-\ket{0}
+\sum_{n=0}^\infty\frac{(2ig_o)^{n}}{(n+2)!}\sum_{m=0}^{n}\bra{B}(Ob_0^-)^{n-m}OQ_BO(b_0^-O)^m\ket{0}\\
&=\frac{1}{4g_o^2}\bra{B}c^-_0Q_Bc_0^-\ket{0}+\frac{1}{2}\bra{B}OQ_BO\ket{0}+\frac{ig_o}{3}\bra{B}\big(OQ_BOb_0^-O+Ob_0^-OQ_BO\big)\ket{0}
+\cdots\,.
\end{align}
Here $g_o$ represents the open string coupling constant.
This expansion form of the BSFT action shares several similar properties with other string field theories.
First note that all the terms are written\footnote
{Though we have used the oscillator formalism of closed string theory to describe BSFT,
we can always go back to expressions written in terms of correlation functions of a conformal field theory.}
in terms of correlation functions of an unperturbed conformal field theory.
The first term is just a constant energy shift and it should correspond to the D-brane tension.
Note that the next term starts from the kinetic term.
Thus, we found that BSFT does not have any open string tadpole around on-shell open string background as desired.
This is not obvious from the conjectured relation \eqref{S=Z} of super BSFT as we shall discuss in the following section.
The kinetic term is quite similar to the one in cubic open string field theory \cite{Witten:1985cc}.
In both theories, the kinetic terms are written by a disk two-point function of boundary vertex operators
$\cal O$ with BRST operator:
\begin{equation}
S_{\rm kin}\sim \vev{{\cal O}Q_B{\cal O}}_{\rm disk}\,.
\end{equation}
The difference is that the positions of two boundary vertex operators $\cal O$ are integrated over the boundary of the disk in the case of BSFT, while they are fixed at some specific points from the beginning in cubic string field theory.
Finally, the interaction part consists of infinitely many terms as in the non-polynomial closed string field theory \cite{nonpoly}
or open superstring field theory \cite{Berkovits:1995ab}.

Obviously, the kinetic term of the above action is invariant under the transformation of $\delta_\Lambda^{(0)} O=[Q_B,\Lambda]$.
The full gauge transformation can be read off from a formal argument of the BV formalism
 even without knowing the BSFT action itself and given by
\begin{equation}
\delta_\Lambda O=[Q_B,\Lambda]
+ig_o\bra{B}\Sym\big[e^{2ig_o\{b_0^-,O\}};\{Q_B,O\},[b_0^-,\Lambda],O_I\big]\ket{0}
\omega^{IJ}O_J.
\end{equation}
Here, $\omega^{IJ}$ is the inverse matrix to $\omega_{IJ}$, which is the components of the two-form $\omega$.
One can directly check the gauge invariance of BSFT by
a straightforward calculation,
\begin{align}
\de_\Lambda S&=\bra{B}\Sym\big[e^{2ig_o\{b_0^-,O\}};\delta_\Lambda O,\{Q_B,O\}\big]\ket{0}\nn\\
&=\bra{B}\Sym\big[e^{2ig_o\{b_0^-,O\}};[Q_B,\Lambda],\{Q_B,O\}\big]\ket{0}\nt
&\quad+ig_o\bra{B}\Sym\big[e^{2ig_o\{b_0^-,O\}};\{Q_B,O\},[b_0^-,\Lambda],\{Q_B,O\}\big]\ket{0}\nn\\
&=\bra{B}\Sym\big[e^{2ig_o\{b_0^-,O\}};[Q_B,\Lambda],\{Q_B,O\}\big]\ket{0}\nt
&\quad+2ig_o\bra{B}\Sym\big[e^{2ig_o\{b_0^-,O\}};[Q_B,\{b_0^-,O\}],\Lambda,\{Q_B,O\}\big]\ket{0}\nn\\
&=-\bra{B}\Sym\big[e^{2ig_o\{b_0^-,O\}};\Lambda,\{Q_B,O\}\big]Q_B\ket{0}\nt
&=0.
\end{align}

Thus, though BSFT appears extraneous to the other covariant string field theories based on
particular conformal field theories and overlapping conditions,
the action of BSFT formally possesses several structures similar to the others.

\section{Revisiting the conjecture $S=Z$}\label{revisit}
As mentioned in the introduction, it is widely believed that the action $S$ of super BSFT is
simply given by the disk partition function $Z$ with boundary perturbations in the case
where matter and ghosts are completely decoupled \cite{KMM2,Marino,Niarchos:2001si,Tseytlin:2000mt}.
In this section, we revisit this well-known conjecture.

For concreteness, let us consider the boundary state for a D$p$-brane.
The GSO-even NS-NS boundary state for a D$p$-brane is given by
a linear combination of two boundary states,
\begin{equation}
{}_{\rm NS}\bra{Dp}={}_{\rm NS}\bra{Bp,+}-{}_{\rm NS}\bra{Bp,-},
\end{equation}
where $\pm$ correspond to different boundary conditions for worldsheet spinor fields as summarized in Appendix \ref{convention}.
However, as we mentioned in section \ref{reformBSFT},
the GSO-even combination is automatically chosen in the action of super BSFT \eqref{act},
due to the GSO-invariance of the closed string state $\newvac$.
Hence we are allowed to focus only on $\bra{Bp,+}$.
Then the BSFT action for a D$p$-brane is given by
\begin{equation}
S=-\frac{i}{4}\bra{Bp,+} e^{\mV}c^-_0Q_Bc_0^-\newvac
-\frac{1}{2} \bra{Bp,+}\Sym\big[e^{\mV};\{Q_B,O\}\big]c_0^-\newvac,\label{act+}
\end{equation}
up to an overall normalization.
Here, we have multiplied the action \eqref{act} by $-i$ in order to make the action real.

When matter and ghosts are completely decoupled,
the string field $O$ is given in the following form:
\begin{equation}
O=\int \frac{d\sigma d\theta}{2\pi}C{\cal V}(\sigma,\theta),
\end{equation}
where $C=c(\si)+\theta\ga(\si)$ is the ghost superfield on the boundary\footnote{
Here, boundary superfields are taken to be the tangential components of the corresponding superfields.
Note that they depend on the boundary condition $\pm$ for worldsheet spinor
fields.
For a precise definition, see Appendix \ref{convention}.}.
The purely matter superfield ${\cal V}$ is expanded as\footnote
{The superscripts indicate the natural picture numbers for the component fields of ${\cal V}$.
However, one should keep in mind that both of them do not have any picture in this context.}
\begin{equation}
{\cal V}={\cal V}^{(-1)}+\theta{\cal V}^{(0)}.\label{matvtx}
\end{equation}
After $\theta$ integration, we have
\begin{equation}
O=\int_0^{2\pi}\frac{d\sigma}{2\pi}(\gamma{\cal V}^{(-1)}(\sigma)-c{\cal V}^{(0)}(\sigma)). 
\end{equation}
The anti-commutator with $b_0^{-}$  in the exponent appearing in the action \eqref{act+} picks the ``zero-picture part''
${\cal V}^{(0)}$ up and the exponential does not depend on ghosts at all:
\begin{equation}
e^{\mV}=\exp\left[\int_0^{2\pi}\frac{d\si}{2\pi}\V^{(0)}\right].
\end{equation}
Therefore, we can freely perform the calculation for ghost parts in this
particular case.

The evaluation of the first term of the action \eqref{act+}
\begin{equation}
S_1\equiv-\frac{i}{4}\bra{Bp,+}e^{\mV}c^-_0Q_Bc_0^-\newvac,
\end{equation}
is straightforward.
Though the BRST operator $Q_B$ appears in this expression, only ghost parts of
it survive due to two $c_0^-$'s
and $S_1$ reduces to the partition function $Z$,
\begin{equation}
S_1=-\bra{Bp,+} e^{\mV}c_0^-\ket\Omega=Z(\la).
\end{equation}

Let us move on to the second term of the action,
\begin{equation}
S_2=-\frac{1}{2} \bra{Bp,+}\Sym\big[e^{\mV};\{Q_B,O\}\big]c_0^-\newvac.\label{S2}
\end{equation}
This term which vanishes for on-shell deformations
potentially represents the correction from the conjectured form $S=Z$.
Defining the following combinations,
\begin{equation}
\be_r^+=\be_r+i\ti\be_{-r},\qquad G^+_r=G_r+i\ti G_{-r},
\end{equation}
which annihilate $\bra{Bp,+}$, the second term becomes
\begin{align}
S_2&=-\frac{i}{2} \bra{Bp,+}\Sym\big[e^{\mV};\{Q_B,O\}\big]c_0^-\beta^+_{1/2}\beta^+_{-1/2}\ket{\Omega}\nn\\
&=-\frac{i}{2} \bra{Bp,+}\Sym\big[e^{\mV};\{G^+_{-1/2},[O,\beta^+_{1/2}]\}\big]c_0^-\ket{\Omega}\nt
&\quad\,-\frac{i}{2} \bra{Bp,+}\Sym\big[e^{\mV};\{G^+_{1/2},[O,\beta^+_{-1/2}]\}\big]c_0^-\ket{\Omega}\nn\\
&\quad\,+\bra{Bp,+}\Sym\Big[e^{\mV};\int_0^{2\pi}\frac{d\si}{2\pi}\V^{(0)}\Big]c_0^-\ket{\Omega}.\label{evS_2}
\end{align}
On the contrary to the anti-commutator with $b_0^-$,
the commutators of $\cal O$ with $\beta^+_{\pm1/2}$ give the ${\cal V}^{(-1)}$
part of the matter superfield with some phases:
\begin{equation}
[O,\beta^+_{\pm1/2}]=i^{1/2}\int \frac{d\sigma}{2\pi}e^{\mp i\sigma/2}{\cal V}^{(-1)}(\sigma).
\end{equation}
The further operation of $G^+_{\mp1/2}$ gives a SUSY transformation as follows:
\begin{align}
&\{G^+_{-1/2},[O,\be^+_{1/2}]\}+\{G^+_{1/2},[O,\be^+_{-1/2}]\}\nt
&=i\int \frac{d\sigma}{2\pi} \left(\oint_\sigma \frac{dw'}{2\pi i}e^{i(w'-\sigma)/2} T_F(w')-\oint_\sigma \frac{d\bar w'}{2\pi i}e^{i(\bar w'-\sigma)/2} \tilde T_F(\bar w')\right){\cal V}^{(-1)}(\sigma)\nt
&\quad+i\int \frac{d\sigma}{2\pi} \left(\oint_\sigma \frac{dw'}{2\pi i}e^{-i(w'-\sigma)/2} T_F(w')-\oint_\sigma \frac{d\bar w'}{2\pi i}e^{-i(\bar w'-\sigma)/2} \tilde T_F(\bar w')\right){\cal V}^{(-1)}(\sigma).\label{wsusy?}
\end{align}
If we restrict the matter superfields to superconformal primaries, which satisfy
\begin{align}
T_F(z){\cal V}^{(-1)}(0)\sim -\frac{1}{z}{\cal V}^{(0)}(0),\label{primary}
\end{align}
the above SUSY transformation \eqref{wsusy?} gives
\begin{equation}
-2i\int \frac{d\sigma}{2\pi} {\cal V}^{(0)}.\label{V0}
\end{equation}
Then, the correction terms \eqref{evS_2} exactly cancel each other.
Thus, we have directly shown that, for purely matter superconformal deformation\footnote
{The previous argument on this conjecture  given in \cite{Niarchos:2001si}, which is based on two decoupled systems,
has also partly relied on superconformal primary fields. On the other hand, the argument given in \cite{Marino} does not
depend on superconformal primary but the proof for \eqref{closedness} and \eqref{V-invariance} is not generally discussed.}, the action \eqref{act} reduces to the partition function $Z$ as conjectured.

The above calculation may also indicate that the conjectured relation $S=Z$ is not always true for operators having
more general OPE than \eqref{primary}.
Actually, one can find some explicit examples in literatures,
which would imply that $S=Z$ is not always valid.
For example, in \cite{Frolov:2001nb, Hashimoto:2004qp}, the authors explicitly calculated correlation functions of some non-primary massive operators in the flat background and found that some one-point functions do not vanish.
These examples suggest that if the action is simply defined by the partition function $Z$, such an action  has a tadpole.
However it is unlikely that such a trivial background is not a solution of the theory.
On the other hand, as shown in the previous section, the general action of (super) BSFT does not have any tadpole around any unperturbed background.
Thus, for operators with non-vanishing one-point functions, we surely need the correction term $S_2$ to cancel the tadpole.

This conclusion is partially satisfactory but there still exists a puzzle.
From the beginning, boundary operators ${\cal O}(\sigma,\theta)$ are assumed
to be superfields. However, we have never used the property of
being superfields except for the argument above,
where we have further restricted ourselves to superconformal primaries.
Thus, throughout the formulation of super BSFT, the superfield formalism
has never shown its importance, though the worldsheet supersymmetry is
known to be necessary to obtain physically reasonable results.
We shall further discuss this puzzle in the next section.

\section{Summary and Discussion}
We have constructed a BV formulation of boundary superstring field theory in terms of boundary states.
We have made a minor modification on the fermionic two-form $\omega$,
which leads to a simpler expression of the two-form when it is written in the closed string Hilbert space.
With this modification,
the proof of the closedness and the invariance of the two-form
under the transformation generated by the fermionic vector $V$,
which are essential to the gauge invariance of BSFT, is much more simplified and transparent.
Furthermore, with the help of the closed string oscillator expression,
we have obtained the general form of the action of super BSFT without any assumption or approximation.
It would be worth while mentioning that
this general action of super BSFT takes exactly the same form as the bosonic one, which enables us to argue formal aspects of the general action of both BSFTs simultaneously as in section \ref{expand}.
In that section, we have derived the expansion form and the gauge transformation of generic BSFT.
We hope that these results help us to understand the relation between BSFT and other formulations of string theory.
Finally, as a special case of the general action, we have revisited the famous conjecture that the action of super BSFT
is simply given by the partition function when matter and ghosts are completely decoupled.
We have directly derived this relation from the general form of the action of super BSFT for superconformal primaries.

In the rest of this section, we argue the puzzle mentioned in the previous section.
The puzzle is that, throughout this paper, the role of the superfield formalism is not clear.
Especially, the proof of closedness and $V$-invariance of the two-form is completely independent from the superfield formalism.
Of course, one would expect that the superfield formalism ensures the existence of the rigid supersymmetry on the worldsheet.
However, the concept of the rigid supersymmetry in the $\sigma$ coordinate is somehow ambiguous due to the anti-periodicity of worldsheet spinor fields in the NS sector.
In order to keep the (anti-)periodicity of superfields, we have to regard the fermionic coordinate $\theta$ as anti-periodic, which implies that $\theta$, and hence, the SUSY transformation parameter implicitly depend on $\sigma$.
One optimistic possibility would be that
the transformation given by \eqref{wsusy?} corresponds to a natural ``rigid'' supersymmetry in the $\sigma$ coordinate
and the superfield formalism here ensures the transformation \eqref{wsusy?} gives the corresponding superpartner \eqref{V0}.
If this is the case, we can always take the relation $S=Z$ as the definition of the super BSFT action as far as
matter and ghosts are decoupled.
However, it also suggests that one-point functions must vanish because
the action of BSFT never has any tadpole as shown in section \ref{expand},
which contradicts the explicit calculations given in \cite{Frolov:2001nb, Hashimoto:2004qp}.
We might have to reconsider the treatment of superfields in the $\sigma$ coordinate more carefully for generic off-shell operators.

\section*{Acknowledgments}
We would like to thank S.~Sugimoto for useful comments and valuable discussions.
We are also grateful to M.~Fukuma, P.~Ho and H.~Nakajima for helpful comments.
Discussions during the YITP workshops YITP-W-07-10 on
``String Phenomenology and Cosmology''
and YITP-W-07-20 on ``30 Years of Mathematical Methods in High Energy Physics''
were useful to complete this work.
This work is supported by Astrophysical Research
Center for the Structure and Evolution of the Cosmos (ARCSEC) and
grant No.~R01-2006-000-10965-0 from the Basic Research Program
through the Korea Science $\&$ Engineering Foundation (A.I.) and
Grants-in-Aid for Scientific Research from the Ministry of Education,
Culture, Sports, Science, and Technology of Japan (S.T.).
\section*{Appendix}
\appendix
\section{Closed String Oscillators, Boundary States and Mode expansions}\label{convention}
Mode expansions:\footnote{
Throughout this paper, we only consider the NS sector. We denote
integer and half-integer modes by the indices $m,n,\cdots$
and $r,s,\cdots$, respectively.}
\begin{align}
\del X^\mu(z)&=-i\sum_{m=-\infty}^{\infty}\frac{\ap^\mu_m}{z^{m+1}},&
\pb X^\mu(\bar z)&=-i\sum_{m=-\infty}^{\infty}\frac{\ti\ap^\mu_m}{\bar z^{m+1}},&
\psi^\mu(z)&=\sum_{r\in {\bf Z}+1/2}\frac{\psi^\mu_r}{z^{r+1/2}},
\end{align}
\begin{align}
b(z)&=\sum_{m=-\infty}^\infty\frac{b_m}{z^{m+2}},&
c(z)&=\sum_{m=-\infty}^\infty\frac{c_m}{z^{m-1}},&
\be(z)&=\sum_{r}\frac{\be_r}{z^{r+3/2}},&
\ga(z)&=\sum_{r}\frac{\ga_r}{z^{r-1/2}}.
\end{align}
Superconformal generators:
\begin{align}
L_m&=\frac12\sum_n\al^\mu_{m-n}\al_{\mu n}
+\frac12\sum_r\left(r-\frac m2\right)\psi^\mu_{m-r}\psi_{\mu r}\nt
&\quad+\sum_n(m+n)b_{m-n}c_n+\frac12\sum_r(m+2r)\be_{m-r}\ga_r-\frac12\de_{m,0}\,,\nt
G_r&=\sum_n\left[\al^\mu_n\psi_{\mu,r-n}
-\frac{2r+n}2\be_{r-n}c_n-2b_n\ga_{r-n}\right].
\end{align}
BRST operator:
\begin{align}
Q_B&=\frac1{2\pi i}\oint(dz\,j_B-d\bar z\,\tilde j_B)\nt
&=\sum_n c_{-n}L_n^\m+\sum_r\ga_{-r}G_r^\m
+\sum_{m,n}\frac{m-n}2 b_{-n-m}c_mc_n\nt
&\quad+\sum_{n,r}\left[\frac{2r-n}2\be_{-m-r}c_m\ga_r-b_{-m}\ga_{m-r}\ga_r\right]
-\frac12c_0+\mbox{(anti-holomorphic part)}.
\end{align}
Ghost zero-modes:
\begin{align}
b_0^+\equiv&b_0+\tilde b_0,&b_0^-\equiv&\half(b_0-\tilde b_0),&
c_0^+\equiv&\half(c_0+\tilde c_0),&c_0^-\equiv&c_0-\tilde c_0.
\end{align}
Normalization:
\begin{equation}
{}_{-1,-1}\bra0c_{-1}\tilde c_{-1}c_0^-c_0^+c_1\tilde c_1\ket0_{-1,-1}=1.
\end{equation}
Boundary state for D$p$-brane:
\begin{align}
\bra{Bp,\pm}={}_{-1,-1}\bra0c_{-1}\tilde c_{-1}c_0^+&
\de^{9-p}(\hat x^i-x^i)
\exp\left[-\sum_{n=1}^\infty \frac{1}{n}\left(\ap_{n}^\mu\tilde{\ap}_{\mu n}
-\ap_{n}^i\tilde{\ap}_{in}\right)
-\sum_{n=1}^\infty(c_{n}\tilde b_{n}+\tilde c_{n} b_{n})\right]\nt
&\times\exp\left[\mp i\sum_{r>0}\left(\psi^\mu_r\ti\psi_{\mu r}
-\psi^i_r\ti\psi_{ir}\right)
\pm i\sum_{r>0}\left(\be_r\ti\ga_r-\ti\be_r\ga_r\right)\right],
\end{align}
where $\mu=0,\ldots,p$ and $i=p+1,\ldots,9$. This satisfies
the following conditions:
\begin{align}
\bra{Bp,\pm}(\ap_{n}^\mu+\tilde \ap_{-n}^\mu)
&=\bra{Bp,\pm}(\ap_n^i-\ti\ap_{-n}^i)
=\bra{Bp,\pm}(c_n+\tilde c_{-n})=\bra{Bp,\pm}(b_n-\tilde b_{-n})=0,\nt
\bra{Bp,\pm}(\psi_r^\mu\mp i\ti\psi_{-r}^\mu)
&=\bra{Bp,\pm}(\psi_r^i\pm i\ti\psi_{-r}^i)
=\bra{Bp,\pm}(\be_r\pm i\ti \be_{-r})
=\bra{Bp,\pm}(\ga_r\pm i\ti \ga_{-r})=0.\label{bc}
\end{align}
In our convention, the boundary of the disk is given by the conditions, $w=\bar w$ and $|z|^2=1$,
in the cylinder coordinate $w=\sigma+i\tau$ and the disk coordinate $z=e^{-i\sigma+\tau}$, respectively.
Boundary operators are written by the tangential components of operators on the boundary.
The mode expansions for matter boundary operators (for the Neumann directions) are given by
\begin{align}
X^\mu(\si)&=x^\mu+i\sum_{m\ne 0}\frac1m
(\al_m^\mu e^{im\si}+\ti\al_m^\mu e^{-im\si}),\\
\psi^\mu_\pm(\si)&=\frac{\psi^\mu(w)\pm\ti \psi^\mu(\bar w)}2
=\frac{i^{-1/2}}2\sum_r (\psi^\mu_r\pm i\ti\psi^\mu_{-r}) e^{ir\si},
\end{align}
and for ghosts,
\begin{align}
b(\si)&=\frac{b(w)+\ti b(\bar w)}2=
-\frac12\sum_n (b_n+\ti b_{-n}) e^{in\si},\\
c(\si)&=\frac{c(w)+\ti c(\bar w)}2=\frac i2\sum_n (c_n-\ti c_{-n}) e^{in\si},\db\\
\be_\pm(\si)&=\frac{\be(w)\pm\ti \be(\bar w)}2
=\frac{i^{-3/2}}2\sum_r (\be_r\mp i\ti\be_{-r}) e^{ir\si},\\
\ga_\pm(\si)&=\frac{\ga(w)\pm\ti \ga(\bar w)}2
=\frac{i^{1/2}}2\sum_r (\ga_r\mp i\ti\ga_{-r}) e^{ir\si},
\end{align}
where the signs $\pm$ represent the possible choices of boundary conditions,
which correspond to the boundary states, $\bra{Bp,\pm}$.
Superfields on the boundary are composed of these tangential components,
\begin{align}
\mb X^\mu_\pm&=X^\mu+2i\theta\psi^\mu_\pm,&
B_\pm&=\be_\pm+\theta b,&C_\pm&=c+\theta\gamma_\pm,
\end{align}
and the supercovariant derivative is $D=\p_\theta+\theta\p_\si$.

\section{BRST operator for boundary fermion}\label{abs}
In section \ref{reformBSFT}, we have proved the $V$-invariance of the
two-form and derived the general action of super BSFT.
The key fact there is that the anti-commutator of the BRST operator $Q_B$ with $b^-_0$ generates the rotation of the whole system.
However, in order to deal with a non-BPS D-brane, we need
to introduce the auxiliary boundary
superfield $\mb\Gamma=\eta+\theta F$, known as the boundary fermion,
to take the GSO-odd sector into account \cite{bdryfermion}.
Therefore, for consistency, we must generalize the BRST operator so that its anti-commutator with $b^-_0$
gives the rotation of the whole sector including the auxiliary field.
In this appendix, we propose such a generalization of the BRST operator.

The kinetic term for the auxiliary boundary superfield $\mb\Gamma$ is given by
\begin{equation}
S_{\mb\Gamma}=-\int\frac{d\si d\theta}{2\pi}\,\mb\Gamma D\mb\Gamma
=-\int\frac{d\si}{2\pi}(\p_\si\eta\,\eta+F^2).\label{Gammakin}
\end{equation}
As an example of the usage of the auxiliary superfield,
the boundary operator for the tachyon is written as
\begin{equation}
O_T=\int\!\frac{d\si d\theta}{2\pi}CT(\mb X)\mb\Gamma
=\int\frac{d\si}{2\pi}\left[
\ga T\eta -cTF-2ic\p_\mu T\psi^\mu\eta\right].
\end{equation}
Recalling that $\eta$ and $F$ are of dimension $0$ and $1/2$, respectively,
we can formally define operators $P_{n}$ and $Q_{r}$ as
\begin{align}
[P_n,\eta]&=ie^{-in\si}\p_\si\eta,& [P_n,F]&=ie^{-in\si}\p_\si F+\frac n2e^{-in\si}F,\nt
\{Q_r,\eta\}&=i^{-3/2}e^{-ir\si}F,&[Q_r,F]&=i^{-3/2}e^{-ir\si}\p_\si\eta.
\end{align}
Then, $P_n$ and $Q_r$ satisfy the super-Virasoro algebra with central
charge zero,
\begin{align}
[P_m,P_n]&=(m-n)P_{m+n}\,,&\{Q_r,Q_s\}&=2P_{r+s}\,,&
[P_n,Q_r]&=\frac{n-2r}2Q_{n+r}\,.\label{alg}
\end{align}
A natural guess for the generalization of the BRST operator would be adding
the following term to the original BRST operator $Q_B$,
\begin{equation}
Q_B'=\sum_n c_nP_{-n}+\sum_r\gamma_rQ_{-r}.\label{modBRST}
\end{equation}
Then, this $Q_B'$ satisfies the following (anti-)commutation relations
as desired,
\begin{equation}
\{Q_B',2b_0^-\}=P_0,\qquad [Q_B,\be_{\pm1/2}^+]=Q_{\pm1/2}.
\end{equation}
The second relation is also essential to ensure the cancellation
of the second term of the general action \eqref{evS_2} for primary operators.
One can also check that this modified BRST operator $Q_B+Q_B'$ is nilpotent:
\begin{equation}
\{Q_B+Q_B',Q_B+Q_B'\}=0.
\end{equation}

As a consistency check of this generalization of the BRST operator,
we demonstrate that the usual tachyon potential can be derived from \eqref{defofS} with this modified BRST operator.
Substituting the constant tachyon profile into \eqref{defofS}, we have
\begin{equation}
dS=dS_1+dS_2+dS_3,
\end{equation}
where
\begin{align}
dS_1&=-i\bra{Bp,+}\eFT
\left(\int\frac{d\si_1}{2\pi}\ga\eta dT\right)
\left(\int\frac{d\si_2}{2\pi}[Q_B,\ga]\eta T\right)\newvac,\db\\
dS_2&=-i\bra{Bp,+}\eFT\left(\int\frac{d\si_1}{2\pi}cF dT\right)
\left(\int\frac{d\si_2}{2\pi} \{Q_B,c\}FT\right)\newvac\nt
&=-\frac12\sum_{n,r}\bra{Bp,+}\eFT\left(\int\frac{d\si_1}{2\pi}cF dT\right)
\left(\int\frac{d\si_2}{2\pi}(\ga_{n-r}\ga_r-\ti\ga_{r-n}\ti\ga_{-r})e^{in\si}FT\right)\newvac,\db\\
dS_3&=-i\bra{Bp,+}\eFT\left(\int\frac{d\si_1}{2\pi}
(\ga\eta-cF) dT\right)
\left(\int\frac{d\si_2}{2\pi}\left(\ga\{Q_B',\eta\}+c[Q_B',F]\right)
T\right)\newvac\nt
&=\bra{Bp,+}\eFT
\left(\int\frac{d\si_1}{2\pi}\ga\eta\,dT\right)
\left[\int\frac{d\si_2}{2\pi}
\Big(\ga\sum_nc_ne^{in\si}-c\sum_ri^{-\frac12}\ga_re^{ir\si}\Big)
\p_\si\eta\,T\right]\newvac\nt
&\quad-i^{\frac32}\sum_r\bra{Bp,+}\eFT\left(\int\frac{d\si_1}{2\pi}cF dT\right)
\left(\int\frac{d\si_2}{2\pi}\ga\ga_re^{ir\si}FT\right)\newvac.
\end{align}
Note that the third term $dS_3$ comes purely from the additional term, $Q_B'$, in the modified BRST operator.
The correlation functions for $\eta$ and $F$ needed to evaluate the
terms above are given by
\begin{align}
\vev{\eta(\si)\eta(\si')}&=\frac\pi 2\ep(\si-\si')
=\frac1{2i}\sum_r\frac{e^{ir(\si-\si')}}r\,,\label{etaeta}\\
\Vev{\eFT F(\si_1)}
&=-\frac12T(\si_1)\,\eTT,\\
\Vev{\eFT F(\si_1)F(\si_2)}
&=\frac14\big[T(\si_1)T(\si_2)-4\pi\de(\si_1-\si_2)\big]\eTT.
\end{align}
However, even before the evaluation, we see that $dS_2$ and the second term of
$dS_3$ cancel out each other by using the boundary condition \eqref{bc}.
Furthermore, the first term of $dS_3$ simply vanishes.
Therefore, the net contribution only comes from $dS_1$, and we have 
\begin{equation}
dS=dS_1=-\frac12TdTe^{-\frac{T^2}4}\db,\label{dS1}
\end{equation}
which leads to the usual tachyon potential $V=e^{-\frac{T^2}4}$.
Note that, though the net contribution comes from the first term, $dS_1$, we also need the cancellation of
the second term, $dS_2$, by the third term, $dS_3$, which originates from the additional term \eqref{modBRST} of our modified BRST operator.
Though we discussed the case of a non-BPS D-brane,
it would be straightforward to extend our argument to multi D-brane systems
(including anti-D-branes) by introducing more than one such auxiliary superfields.


\begin{thebibliography}{99}
\bibitem{Witten:1992qy}
E.~Witten,
``On background independent open string field theory,''
Phys.\ Rev.\ D {\bf 46}, 5467 (1992)
[arXiv:hep-th/9208027].
%%CITATION = HEP-TH 9208027;%%

\bibitem{Witten:1992cr}
E.~Witten,
``Some computations in background independent off-shell string theory,''
Phys.\ Rev.\ D {\bf 47}, 3405 (1993)
[arXiv:hep-th/9210065].
%%CITATION = HEP-TH 9210065;%%

\bibitem{Witten:1985cc}
  E.~Witten,
``Noncommutative Geometry And String Field Theory,''
  Nucl.\ Phys.\ B {\bf 268}, 253 (1986).
  %%CITATION = NUPHA,B268,253;%%

\bibitem{Berkovits:1995ab}
  N.~Berkovits,
``SuperPoincare invariant superstring field theory,''
  Nucl.\ Phys.\  B {\bf 450}, 90 (1995)
  [Erratum-ibid.\  B {\bf 459}, 439 (1996)]
  [arXiv:hep-th/9503099].
  %%CITATION = NUPHA,B450,90;%%

\bibitem{Sen:2004nf}
  A.~Sen,
``Tachyon dynamics in open string theory,''
  Int.\ J.\ Mod.\ Phys.\ A {\bf 20}, 5513 (2005)
  [arXiv:hep-th/0410103].
  %%CITATION = HEP-TH 0410103;%%

\bibitem{Gerasimov:2000zp}
  A.~A.~Gerasimov and S.~L.~Shatashvili,
``On exact tachyon potential in open string field theory,''
  JHEP {\bf 0010}, 034 (2000)
  [arXiv:hep-th/0009103].
  %%CITATION = HEP-TH 0009103;%%

\bibitem{KMM}
  D.~Kutasov, M.~Marino and G.~W.~Moore,
``Some exact results on tachyon condensation in string field theory,''
  JHEP {\bf 0010}, 045 (2000)
  [arXiv:hep-th/0009148].
  %%CITATION = HEP-TH 0009148;%%

\bibitem{KMM2}
  D.~Kutasov, M.~Marino and G.~W.~Moore,
``Remarks on tachyon condensation in superstring field theory,''
  arXiv:hep-th/0010108.
  %%CITATION = HEP-TH/0010108;%%

\bibitem{Shatashvili:1993ps}
  S.~L.~Shatashvili,
``Comment on the background independent open string theory,''
  Phys.\ Lett.\ B {\bf 311}, 83 (1993)
  [arXiv:hep-th/9303143];
  %%CITATION = HEP-TH 9303143;%%
``On the problems with background independence in string theory,''
  Alg.\ Anal.\  {\bf 6}, 215 (1994)
  [arXiv:hep-th/9311177].
  %%CITATION = HEP-TH 9311177;%%

\bibitem{oldS=Z}
A.~A.~Tseytlin,
``Vector Field Effective Action In The Open Superstring Theory,''
Nucl.\ Phys.\  B {\bf 276}, 391 (1986)
[Erratum-ibid.\  B {\bf 291}, 876 (1987)];
%%CITATION = NUPHA,B276,391;%%
``Renormalization of Mobius Infinities and Partition Function Representation
for String Theory Effective Action,''
Phys.\ Lett.\  B {\bf 202}, 81 (1988);
%%CITATION = PHLTA,B202,81;%%

O.~D.~Andreev and A.~A.~Tseytlin,
``Partition Function Representation for the Open Superstring Effective
Action: Cancellation of Mobius Infinities and Derivative Corrections to
Born-Infeld Lagrangian,''
Nucl.\ Phys.\  B {\bf 311}, 205 (1988).
%%CITATION = NUPHA,B311,205;%%

\bibitem{DDbar}
  P.~Kraus and F.~Larsen,
``Boundary string field theory of the DD-bar system,''
  Phys.\ Rev.\  D {\bf 63}, 106004 (2001)
  [arXiv:hep-th/0012198];
  %%CITATION = PHRVA,D63,106004;%%

T.~Takayanagi, S.~Terashima and T.~Uesugi,
``Brane-antibrane action from boundary string field theory,''
  JHEP {\bf 0103}, 019 (2001)
  [arXiv:hep-th/0012210].
  %%CITATION = JHEPA,0103,019;%%

\bibitem{Marino}
  M.~Marino,
``On the BV formulation of boundary superstring field theory,''
  JHEP {\bf 0106}, 059 (2001)
  [arXiv:hep-th/0103089].
  %%CITATION = HEP-TH 0103089;%%

\bibitem{Niarchos:2001si}
  V.~Niarchos and N.~Prezas,
``Boundary superstring field theory,''
  Nucl.\ Phys.\ B {\bf 619}, 51 (2001)
  [arXiv:hep-th/0103102].
%%CITATION = HEP-TH 0103102;%%

\bibitem{Teraguchi:2006tb}
  S.~Teraguchi,
``Reformulation of boundary string field theory in terms of boundary state,''
  JHEP {\bf 0702}, 017 (2007)
  [arXiv:hep-th/0610171].
%%CITATION = JHEPA,0702,017;%%

\bibitem{FMS}
  D.~Friedan, E.~J.~Martinec and S.~H.~Shenker,
``Conformal Invariance, Supersymmetry And String Theory,''
  Nucl.\ Phys.\  B {\bf 271}, 93 (1986).
  %%CITATION = NUPHA,B271,93;%%

\bibitem{DSTEP}
C.~R.~Preitschopf, C.~B.~Thorn and S.~A.~Yost,
``SUPERSTRING FIELD THEORY,''
  Nucl.\ Phys.\  B {\bf 337}, 363 (1990);
  %%CITATION = NUPHA,B337,363;%%

I.~Y.~Arefeva, P.~B.~Medvedev and A.~P.~Zubarev,
``New Representation For String Field Solves The Consistence Problem For Open
Superstring Field,''
Nucl.\ Phys.\  B {\bf 341}, 464 (1990);
%%CITATION = NUPHA,B341,464;%%
``Background Formalism For Superstring Field Theory,''
Phys.\ Lett.\  B {\bf 240}, 356 (1990).
%%CITATION = PHLTA,B240,356;%%

\bibitem{SuperCSFT}
E.~Witten,
``Interacting Field Theory of Open Superstrings,''
Nucl.\ Phys.\  B {\bf 276}, 291 (1986).
%%CITATION = NUPHA,B276,291;%%

\bibitem{Kmtx}
T.~Asakawa, S.~Sugimoto and S.~Terashima,
``D-branes, matrix theory and K-homology,''
  JHEP {\bf 0203}, 034 (2002)
  [arXiv:hep-th/0108085];
  %%CITATION = JHEPA,0203,034;%%
``Exact description of D-branes via tachyon condensation,''
  JHEP {\bf 0302}, 011 (2003)
  [arXiv:hep-th/0212188].
  %%CITATION = JHEPA,0302,011;%%

\bibitem{bdryfermion}
E.~Witten,
``D-branes and K-theory,''
JHEP {\bf 9812}, 019 (1998)
[arXiv:hep-th/9810188];
%%CITATION = JHEPA,9812,019;%%

J.~A.~Harvey, D.~Kutasov and E.~J.~Martinec,
``On the relevance of tachyons,''
arXiv:hep-th/0003101.
%%CITATION = HEP-TH/0003101;%%

\bibitem{nonpoly}
B.~Zwiebach,
``Closed string field theory: Quantum action and the B-V master equation,''
Nucl.\ Phys.\ B {\bf 390}, 33 (1993)
[arXiv:hep-th/9206084].
%%CITATION = HEP-TH 9206084;%%

\bibitem{Tseytlin:2000mt}
A.~A.~Tseytlin,
``Sigma model approach to string theory effective actions with tachyons,''
J.\ Math.\ Phys.\  {\bf 42}, 2854 (2001)
[arXiv:hep-th/0011033].
%%CITATION = JMAPA,42,2854;%%

\bibitem{Frolov:2001nb}
S.~A.~Frolov,
``On off-shell structure of open string sigma model,''
JHEP {\bf 0108}, 020 (2001)
[arXiv:hep-th/0104042].
%%CITATION = JHEPA,0108,020;%%

\bibitem{Hashimoto:2004qp}
K.~Hashimoto and S.~Terashima,
``Boundary string field theory as a field theory: Mass spectrum and
interaction,''
JHEP {\bf 0410}, 040 (2004)
[arXiv:hep-th/0408094].
%%CITATION = HEP-TH 0408094;%%


\end{thebibliography}
\end{document}